\documentstyle[11pt]{article}

\def\etal{{\it et al.}}

\def\gtwid{\mathrel{\raise.3ex\hbox{$>$\kern-.75em\lower1ex\hbox{$\sim$}}}}
\def\ltwid{\mathrel{\raise.3ex\hbox{$<$\kern-.75em\lower1ex\hbox{$\sim$}}}}
\begin{document}
\begin{titlepage}
\begin{flushright}
OSU-TA-13/95\\
BA-95-25\\
astro-ph/9507003
\end{flushright}
\begin{center}
\Large
{\bf Gauge Invariant Density and Temperature \\
        Perturbations in the Quasi--Newtonian\\
            Formulation  }
\\
\vspace{.6cm}
\normalsize
\large
R.K. Schaefer, \\
{\em Bartol Research Institute,\\
University of Delaware,\\
Newark, DE  19716}\\
\vspace{.6cm}
Andrew A.\ de Laix,\\
{\em Physics Department, \\
Ohio State University\\
Columbus, OH 432310}
\end{center}
\normalsize
\vspace{.2 cm}

\begin{abstract}
    We here give an improved formalism for calculating the evolution of
density fluctuations and temperature perturbations in flat universes.  Our
equations are general enough to treat the perturbations in collisionless
relics like massive neutrinos.  We
find this formulation to be simpler to use than gauge dependent and other
gauge--invariant formalisms.  We show how to
calculate temperature fluctuations (including multipole moments) and transfer
functions, including the case of collisionless relics like massive
neutrinos.  We call this formalism ``quasi-Newtonian" because the
equations for the potential and cold matter fluctuation evolution have the same
form as the Newtonian gravitational equations in an expanding space.  The
density fluctuation variable also has the same form inside and outside of the
horizon which allows the initial conditions to be specified in a simple
intuitive way.  Our sample calculations demonstrate how to use these equations
in cosmological models which have hot, cold, and mixed dark matter and
adiabatic (isentropic) or isocurvature modes.  We also give an
approximation may be used to get transfer functions quickly.

\end{abstract}
\begin{flushleft}
{\sl Subject Headings:} cosmology: theory - dark matter - cosmic microwave
background
\end{flushleft}
\end{titlepage}

\section{Introduction}

   Finding the correct model for the formation of structure in the universe is
currently a subject of intense study.  It is generally believed that cosmic
structures are the result of the gravitational amplification of initially
small density fluctuations in the early universe.  There are candidate
mechanisms for producing the primordial fluctuations, {\it e.g.}, inflation and
topological defects, which must be tested against observations of the current
universe.  To test these primordial mechanisms, one has to evolve the
perturbations up to the present day.  This task is complicated by the fact
that the amount and composition of dark matter is not known, so potentially
a large number of possibilities may exist which need to be checked against
observations.

   The two main pieces of information required for testing theories are the
predicted power spectra of the temperature and matter fluctuations.  Various
prescriptions for how to calculate these exist in the literature (Kodama \&
Sasaki, 1984: Sasaki \& Gouda, 1986; Bond \& Efstathiou, 1987; Mukhanov,
Feldman, \& Brandenberger, 1987; Durrer \& Straumann, 1988; Stompor, 1994).
The problem with these treatments is that either they do not span the current
range of models or they that use variables which are unsatisfying in that
the physical nature of the initial conditions in is not transparent.

   Many of the treatments do not include equations for how to deal with
the free streaming of massive collisionless relics such as few eV mass
neutrinos, an element which is indicated in critical density universe
models based on inflation (Schaefer \& Shafi, 1994; Pogosyan \&
Starobinsky, 1994; Liddle \& Lyth, 1994).  Some gauge-invariant treatments
which include a treatment
of collisionless relics (Durrer \& Straumann, 1988; Stompor, 1994) use a
generalized density perturbation variable (there are several reasonable
candidates) which has a complicated behavior
outside of the horizon and thus makes it inconvenient for specifying initial
conditions.  Here we present a unified treatment of the evolution of matter and
temperature perturbations, which include free streaming massive neutrinos, and
use a density perturbation variable which is most natural to use from the point
of view of matter perturbations.  We also have attempted to put this into a
form which can be easily used by someone to get results without the
need to slog through the subtleties of the general relativistic underpinnings
of the formalism.

For the sake of brevity we have only treated flat universes here.  We expect
to follow this paper with another which covers universes with curvature.

   We begin in section 2 by constructing a gauge-invariant definition of the
distribution function, in a slightly different form than that of Durrer \&
Straumann, (1988).  We then recast the generally covariant Boltzmann equation
for this distribution function into a more useful form, and this is the basis
for the rest of the paper.  We also derive fluid equations for the first
three moments of the distribution function.  In section 3 we discuss how to
treat relativistic collisionless particles, {\it e.g.} massless neutrinos and
decoupled photons, using only fluid variable
equations.  Section 4 explains how to
derive initial conditions.  In section 5 we discuss how to
treat the baryon-photon fluids which are coupled early on, and then show how to
calculate the present day temperature anisotropies.   Section 6 gives
examples of numerical
implementation of these equations, and some details about the numerical
computations.  We end with a summary of our main results.
Lastly, there is an appendix which gives a quick way to calculate transfer
functions for some models.

\section{General Gauge-Invariant Evolution Equations for Collisionless Relics}

    Collisionless massive relic particles cannot be adequately described in a
fluid
formulation of the gravitation equations, (unless the particles are ``cold" on
the scales of interest, {\it i.e.} their thermal energy is much less than the
gravitational potential of a density perturbation.  One must go back to the
Boltzmann equation.  Since we are working in a regime in which general
relativistic effects are important, the best place to start is with
the generally
covariant Boltzmann equation.  This equation has been derived and discussed in
many places (see e.g., Lindquist, 1966; Ehlers, 1969; Stewart, 1971; Kodama
\& Sasaki, 1984; Durrer \& Straumann, 1988), and we need not repeat the
details here.  [An interesting approach to solving this equation in power
series form has also been explored (Rebhan \& Schwartz, 1994).]  We will
only sketch
the pertinent parts of the derivation of the perturbed Boltzmann equation.
Small perturbations of the distribution function of the particles are not in
general gauge invariant.  We will remove the gauge dependent part
from the distribution function.  Next we expand the gauge-invariant
distribution function in terms of angular moments, the first three of which
are associated with gauge invariant fluid variables: the density, velocity,
entropy, and anisotropic pressure perturbations.  We will then discuss the
correspondence of the initial values
of these moments to the initial fluid variable values.  We end this
section with some remarks on the physical interpretation of the equations.

\subsection{ The Generally Covariant Boltzmann Equation}

The non--relativistic distribution function specifies the number of
particles at time $t$ with velocity between ${\vec v}$ and ${\vec v} +
d{\vec v} $
and position between ${\vec x}$ and ${\vec x} + d{\vec x}$.  Since we intend to
describe the behavior of relativistic particles in a perturbed spacetime,
we need to use a
generally covariant formulation, so we must take care that our definition of
the distribution function is
also generally covariant.  For the setup of the Boltzmann equation, we
follow the treatment of Kodama \& Sasaki, (1984).  We will also borrow from
Durrer \& Straumann, (1988).

     In the general relativistic version of kinetic theory, we use
the invariant volume elements to define the distribution function.  Here we
will need the space-like covariant version of the volume element
$\sigma_\alpha$:
\begin{equation} \sigma_\alpha = -{1 \over 3!}u_\alpha u^\mu
\epsilon_{\mu\nu\lambda
\sigma}dx^\nu dx^\lambda dx^\sigma \end{equation}
where $u^\mu$ is an arbitrary time like vector field and $\epsilon_{\mu\nu
\lambda\sigma}$ is the totally antisymmetric tensor with $\epsilon_{0123}
= \surd (- \hbox{det} g_{\mu \nu})$.  $g_{\mu \nu}$ is the metric.  Greek
indices run from 0 to 3, while latin indices i through m will run from
1 to 3.  The other latin indices will be used to distinguish the different
component fluids.  The
invariant momentum space volume element for a particle with mass $m$ is
$\pi_q$
\begin{equation}
\label{momentumve}
\pi_q = {2 \over 4!}\theta(q^0)\delta(g_{\mu \nu}q^\mu q^\nu +m^2)
\epsilon_{\mu\nu\lambda\sigma}dq^\mu dq^\nu dq^\lambda dq^\sigma.
\end{equation}
Here $q^\mu$ is the contravariant momentum, the $\delta$ function enforces
the relation among energy mass and momentum
(keeps the particle momenta ``on the mass shell"), and the Heavyside ($\theta$)
function restricts us to the regime of positive energies for the particles.
The number of particles crossing a unit hypersurface element
$\sigma_\alpha$ within a momentum space volume $\pi_q$ is given by
\begin{equation} dn = f(x^\alpha, q^\beta)q^\mu \sigma_\mu \pi_q.
\end{equation}
Conservation of $q^\mu \sigma_\mu \pi_q$ along a particle trajectory is
guaranteed by the Liouville theorem.  Any change in $dn$ due to interparticle
collisions can be represented by
\begin{equation}C(f)q^\mu \sigma_\mu \pi_q.
\end{equation}
The relativistic Boltzmann equation becomes
\begin{equation}{\cal L}(f) = C(f).
\end{equation}
where ${\cal L}(f)$ is the linear operator which is the total derivative
of the distribution function with
respect to an affine parameter $\lambda$ which traces geodesics.  In
non-relativistic theory $\lambda$ would be identified with the time
variable. In our relativistic version, $\cal L$ is
\begin{equation}
{\cal L} = {d\over d\lambda} = q^\mu{d\over dx^\mu} +{dq^\mu \over d\lambda}
{d\over dq^\mu}
\end{equation}
where the momentum is defined, as usual, by
\begin{equation} q^\mu = {dx^\mu \over d\lambda},
\end{equation}
and the particles follow geodesics which satisfy
\begin{equation}{dq^\mu \over d\lambda} = - \Gamma^\mu_{\alpha \beta}q^\alpha
q^\beta
\end{equation}
The $\Gamma^\mu_{\alpha \beta}$ are the metric connection coefficients.
Since we are initially interested in collisionless particles we will set
$C(f) = 0$.  In section 5 we will revisit the collision term for baryon--photon
scattering.

    Eventually, we will need to relate this microscopic distribution function
to macroscopic quantities like the density and pressure.  This is done by
taking momentum averages.  The energy-momentum tensor can be obtained from
the formula
\begin{equation}T^{\mu \nu} = \int q^\mu q^\nu f \pi_q
\end{equation}
In order to evaluate these integrals and solve the Boltzmann equation, one
usually introduces a tetrad (or vierbein) frame which is convenient for the
purpose, usually the one which looks like flat space so that dot products are
vectors contracted with the Minkowski metric.  Kodama \& Sasaki (1984)
introduce a set of tetrads on which the momentum is physical and redshifts with
the universal expansion.  Bond \& Szalay (1983) and Durrer \& Straumann
(1988) use a tetrad frame on which the momentum
are comoving in flat expanding space, which has the advantage that the
spatial momentum components are constant in time.  Here we will follow the
latter convention, and refer interested readers
to the original works for more details about the Boltzmann equation.

One can
define in a covariant way the one-particle distribution function $f$ as a
scalar function of (conformal) time $\tau$, the comoving three-space
coordinates $x^i$, and the three-momentum $\tilde{p}^i$ (for particles on
the mass
shell).  We want to consider perturbations on a homogeneous and isotropic
background universe, so we split the distribution function into background $+$
perturbation.  The spatial dependence of the perturbations can be expanded in
terms of harmonic functions $Y(k^i,x^i)$, which for flat space (e.g.,
critical density universes) is like Fourier transforming the equations, as
$Y=e^{ik^jx_j}$. We then split the distribution function into
\begin{equation}
\label{fndeltaf}
f = f^0 + (\delta f) Y
\end{equation}
where the perturbation $\delta f$ depends on the Fourier wavevector $k^i$, the
time $\tau$ and the three momentum.  We also perturb the metric $g_{\mu\nu}$ as
\begin{equation}
g_{\mu\nu} = a^{-2}\left[\eta_{\mu\nu} + h_{\mu\nu}\right]
\end{equation}
where $a$ is the time dependent ``scale factor", and $\eta_{\mu\nu}$ is the
Minkowski metric.  The scale factor satisfies the equation
\begin{equation}
\label{adotovera}
\left( {\dot{a}\over a}\right)^2 = {8 \pi G\over 3} \rho a^2 + {\Lambda
a^2\over 3},
\end{equation}
where $G$ is Newton's gravitation constant, $\rho$ is the
total density of mass in the universe, and $\Lambda$ is the cosmological
constant.  In the above equation and throughout
the paper, the dot ($\cdot$) over a variable denotes a derivative with respect
to conformal time $\tau$.  The conformal time is more convenient for doing
these calculations than the real time $t$.  The conformal time is, in fact,
the co-moving horizon size.  This property is useful for relating the
scale of different effects to the scale of structures in the present universe.

  The metric perturbation $h_{\mu\nu}$ can also be expanded in
terms of harmonic functions:
\begin{eqnarray}
h_{00} &=& -2 A(k,\tau) Y\nonumber \\
h_{0j} &=& i{k_j\over k} B(k,\tau) Y \nonumber \\
h_{ij} &=& -2 H_L(k,\tau)Y \delta_{ij} - 2 H_T(k,\tau)
\left[{\delta_{ij}\over 3} -{k_i k_j\over k^2}\right] Y.
\end{eqnarray}

   The energy momentum tensor $T^{\mu\nu}$ is also perturbed.  We can write the
perturbations in terms of the usual (gauge dependent) fluid perturbations
($\delta$ for the density perturbation, $v_f$ for the fluid velocity
perturbation, and $\pi_L$ or $\pi_T$ for the isotropic or anisotropic pressure
perturbations) and $Y$:
\begin{eqnarray}
T_{0}^0 &=& -\rho\left[ 1 +  \delta(k,\tau) Y\right]\nonumber \\
T_{0}^j &=& -i{k_j\over k}(\rho+p) v_f(k,\tau) Y \nonumber \\
T_{j}^i &=& -p\left[ \delta^i_j + \pi_L(k, \tau) Y \delta^i_j + \pi_T(k, \tau)
\left({\delta_{j}^i\over 3}-{k_i k_j\over k^2}\right) Y \right].
\end{eqnarray}

   Following Durrer \& Straumann (1988), we will write the Boltzmann equation
(in the tetrad frame) in terms of the comoving momentum $v = \tilde{p}/T$,
where
$\tilde{p}$ is the magnitude of the three vector momentum and T is the
temperature parameter of the
particles.  The Boltzmann equation with a more physical definition of the
momentum which redshifts with time, such as found in (Kodama \& Sasaki, 1984;
Schaefer, 1991) while being more physically accurate is more complicated to
solve numerically, so we avoid it.  The particle energy variable $q$ (following
Durrer \& Straumann, 1988) associated with the comoving momentum is
\begin{equation}
q = \sqrt{v^2 + {m^2\over T^2}}
\end{equation}
Of course we are using units for which $c=k_B=1$.  The physical momentum
$\tilde{p}$ and the temperature $T$ both decrease with time as
$1/a$, so that $v$ is time independent, but the energy variable is not.
In this notation the density and pressure are given by
\begin{equation}
\rho = T^4 \int d\Omega dv v^2 q f,
\end{equation}
and
\begin{equation}
p = T^4 {1\over 3}\int d\Omega dv {v^4 \over q} f.
\end{equation}

Since on average the universe is homogeneous and
isotropic, the background distribution function must also be homogeneous and
isotropic and is fixed when (and if) the particles were in thermal
equilibrium in the very early universe:
\begin{equation}
f^0 = {1\over (2\pi{\mathchar'26\mskip-9mu h})^3}{1\over 1 \pm e^v}
\end{equation}
where the $+$ is for fermions and the $-$ is for bosons, and
${\mathchar'26\mskip-9mu h}$ is Planck's constant divided by $2\pi$.

\subsection{ The Gauge--Invariant Perturbed Boltzmann Equation}

The Boltzmann equation for the perturbation to $f$ in eq. (\ref{fndeltaf}) is
then
\begin{eqnarray}
({\delta f})^{\bf \cdot} &+&i k \mu {v\over q} \delta f = \nonumber \\
& &{df^0\over dv}\left[ ik\mu q A + v(\dot{H_L} + {1\over3}\dot{H_T} )
-  \mu^2 v (\dot{H_T} - kB) \right].
\end{eqnarray}
The perturbation variables ($\delta f$, $A$, $B$, and $H_L$) are gauge
dependent, and so are not the most physically relevant variables to work
with.  Durrer
\& Straumann (1988) showed how to construct a gauge invariant version
of the distribution function perturbation.  The full procedure for doing this
is given there.  Operationally, all we need to know is that we must add
something to $\delta f$ so that the appropriate integrals of $\delta f$ yield
the gauge--invariant density and velocity perturbations.  The construction in
Durrer \& Straumann (1988) was not unique, as one could
subtract off the gauge dependence in other ways.  In fact they
chose the combination such that when one converts the distribution function
perturbation to its corresponding fluid variable density perturbation by
integrating over momentum, one would get the density perturbation variable
called
$\Delta_g$ by Kodama \& Sasaki, (1984), $\epsilon_g$ by Bardeen, (1980), and
$\Delta$ by Mukhanov, Feldman, \& Brandenberger, (1991).
\begin{equation}
\label{deltag}
\Delta_g = \delta + 3\left( 1 + w\right) \left( H_L + {1\over 3} H_T \right),
\end{equation}
where $w\equiv p/\rho$.  Instead we would
like to make another choice for a gauge--invariant distribution function in
which the corresponding gauge invariant density perturbation is the variable
called $\Delta$ by Kodama \& Sasaki (or $\Delta_{ca}$ for a particular
component--note that Kodama \& Sasaki also define a component perturbation
variable called $\Delta_a$, which is {\em not} the one we are
interested in here.)
\begin{equation}
\label{deltadef}
\Delta = \delta + 3 ( 1 + w ) {\dot{a}\over a} ( v_f - B)
\end{equation}
This variable is $\epsilon_m$ in Bardeen's notation.  It is the
density perturbation on the spacelike hypersurfaces which are the local matter
rest frame, and hence is
the most natural variable to use from the point of the matter fluctuations.
Using this variable, the fluid equations most resemble their Newtonian
counterparts for a fluid under the action of a gravitational perturbation in
expanding coordinates.  Also, this is the density perturbation which
Hu \& Sugiyama (1995) used in order to untangle the various effects
which contribute to the cosmic microwave anisotropy.  We will comment on
this aspect in more detail later.  We now set about rewriting the
Liouville equation in a
gauge-invariant way, and show explicitly how to use it to calculate
the growth of density perturbations in critical density universes.

The gauge invariant generalization ${\cal F}$ of the distribution function
perturbation $\delta f$ is given here by
\begin{equation}
\label{F}
{\cal F} = \delta f - {df\over dv}\left[{\dot{a}\over a} {v\over k}(v_f-B)
+ i\mu {q\over k}(\dot{H_T} - kB) \right].
\end{equation}
Plugging this form into the Boltzmann equation, we get (after some algebraic
manipulation)
\begin{eqnarray}
\label{Fdot}
 \dot{\cal F} &+& ik\mu {v\over q} {\cal F} + {df\over dv}\left[{v\over 1 +w}
{\dot{a}\over a}\left(c_s^2 \Delta +w\Gamma -{2\over3}w\Pi \right)\right]
 \nonumber \\
& &+i\mu {df\over dv}\left[ {3q\over2 k}\left[{\dot{a}\over a} \right]^2
(\Delta + 2 w \Pi) + {\dot{a}\over a}{v^2\over q}V \right]=0
\end{eqnarray}
where all the terms are now manifestly gauge invariant.  In the above $V$ is
now the gauge-invariant fluid velocity perturbation, $\Gamma$ is the
perturbation in the entropy of the system, and $\Pi$ is the amplitude of the
anisotropic pressure perturbation.  The ``sound velocity" $c_s^2$ is defined as
$c_s^2  \equiv \dot{p} / \dot{\rho}$.
Note that there is really only one obvious
choice for the gauge--invariant velocity perturbation $V= v_f - \dot{H}_T/k$
(Bardeen, 1980; Kodama \& Sasaki, 1984; Lyth \& Bruni, 1994).  The other
terms $\Gamma \equiv \pi_L - c_s^2 \delta /w$ and $\Pi \equiv \pi_T$ are
already gauge--invariant.

\subsection{Angular Moments of the Distribution Function}

   The equation for ${\cal F}$ can be solved directly, and then one has to do a
double integral over $\mu$ and $v$ to get quantities like the density
perturbation and velocity perturbation as in (Durrer, 1989).   However, one can
instead expand the angular dependence of the distribution function perturbation
in orthogonal polynomials, namely Legendre polynomials $P_\ell(\mu)$
(Valdarnini \& Bonometto, 1985; Schaefer, 1991) as
\begin{equation}
\label{Ftosig}
{\cal F} = {1\over 4 \pi}\sum_{\ell = 0}^{\infty} (2\ell +1) i^\ell P_\ell(\mu)
\sigma_\ell(k,\tau)
\end{equation}
and plugging this form into the Boltzmann equation above.  Expanding the
product $(2\ell +1) \mu P_\ell(\mu) = \ell P_{\ell-1} + (\ell+1) P_{\ell+1}$,
resumming the series, and projecting out the coefficients of $P_\ell$, we get
the following set of equations for the $\sigma_\ell$

%
%
\begin{eqnarray}
\label{sigdot}
\dot{\sigma_\ell} &+& k {v\over q}\left[{\ell \over 2 \ell + 1}\sigma_{\ell-1}
-{\ell +1\over 2 \ell+1}\sigma_{\ell+1}\right] = \nonumber \\
&-&4\pi {df^0\over dv}\bigg[\delta_{\ell 0}{v\over 1+w} {\dot{a}\over a}(c_s^2
\Delta + w \Gamma - {2\over 3}w\Pi)  \nonumber \\
& & +{1\over 3}\delta_{\ell 1}
\left({3q\over 2 k}\left({\dot{a}\over a}\right)^2(\Delta + 2 w \Pi)  +
{v^2\over
q}{\dot{a}\over a} V  \right)\bigg]
\end{eqnarray}
where $\delta_{ij}$ is the Kronecker delta and $\sigma_{-1}=0$.  The fluid
variables, shown here for component $a$, can be obtained now from
integrals over only the magnitude of the momentum $v$:
\begin{equation}
\label{del}
\Delta_{ca} =  {T_a^4\over \rho_a}\int_0^\infty dv v^2 q\sigma_0
\end{equation}
\begin{equation}
\label{V}
V_a = -{T_a^4\over \rho_a + p_a}\int_0^\infty dv v^3\sigma_1
\end{equation}
\begin{equation}
\label{Pi}
w_a \Pi_a = {T_a^4\over \rho_a }\int_0^\infty dv {v^4\over q}\sigma_2
\end{equation}
\begin{equation}
\label{Gamma}
w_a \Gamma_a = {T_a^4\over \rho_a }\int_0^\infty dv \left[{v^4\over 3 q}
- c_a^2 v^2 q\right] \sigma_0
\end{equation}
The density perturbation is $\Delta_{ca}$, the velocity perturbation is $V_a$,
$\Pi_a$ is the anisotropic pressure perturbation, and $\Gamma_a$ is the
internal entropy perturbation, which becomes non-zero when density and pressure
perturbations get out of phase.
One can derive from the above equations the values for the total
perturbations for a multi--component system, finding
%
%
\begin{eqnarray}
\label{totalfluid}
\Delta  &=&  \rho ^{-1}\sum_{}^{}\rho _{a}\Delta _{ca} \\
V &=& (\rho + p)^{-1}\sum_{}^{}(\rho _{a}+p_{a}) V_{a} \\
\Pi  &=& p^{-1}\sum_{}^{}p_{a}\Pi _{a} \\
\label{totalfluidend}
\Gamma  &=& p^{-1}\sum_{}^{}\left[ p_{a}\Gamma _{a} + (c_{a}^{2} -
c_{s}^{2})\rho _{a}\Delta _{ca} \right].
\end{eqnarray}
Using the above relations, the expression in equations (\ref{Fdot} \&
\ref{sigdot}) which is common to all the component fluid equations can be
simply expressed as
\begin{equation}
\label{pressr}
c_s^2 \Delta + w \Gamma -{2\over3} w\Pi = \rho ^{-1}\sum_{}^{}\rho _{a}
( c_a^2 \Delta _{ca} +w_a \Gamma_a -{2\over3} w_a\Pi_a).
\end{equation}
One disadvantage of this formalism is that one must now solve a possibly
large number of coupled $\sigma_\ell$ equations.   However, one can regulate
the number of $\sigma_\ell$ calculated to get an accurate answer, which can be
estimated ahead of time.  Specifically, one would like to ensure that the
logarithmic derivative $d \log \sigma_{\ell}/d \log \tau $ is small.
Roughly, this condition is met when $k \tau (T/m)/\ell \ll 1$ for a
massive particle (if the particle is massless, just ignore the $T/m$
term) over the entire range of integration.  For a particle which becomes
non-relativistic prior to the epoch of matter/radiation equality, one can
easily show that the constraint is most stringent when evaluated at
equality, thus one should satisfy the condition $k \tau (T/m)/\ell|_{eq}
\ll 1$ to ensure an accurate integration.  Should one be interested in an
extremely fast integration in which lower accuracy can be tolerated, one can
use the abbreviated fluid version of these equations as described in the
Appendix.

One can then integrate the moments [using the procedure as given, for example
in Schaefer (1991)] to recover the fluid equations given in other work
(Bardeen, 1980; Kodama \& Sasaki, 1984).  For example, multiplying the
equation for $\sigma_0$ by $T_a^4 v^2 q/\rho_a$ and integrating over $v$ we get
\begin{eqnarray}
\label{deldot}
\dot{\Delta}_{ca} &+& 3 {\dot{a}\over a}(c_a^2 - w_a)\Delta_{ca} = - 3
{\dot{a}\over a} w_a \Gamma_a \nonumber\\
& & -k(1+w_a)V_a +3 {\dot{a}\over a} {1+ w_a\over 1+w}
\left[ c_s^2 \Delta + w \Gamma - {2\over 3}w \Pi \right].
\end{eqnarray}
 Similarly we obtain from the equation for $\sigma_1$
\begin{eqnarray}
\label{Vdot}
\dot{V}_a &+&  {\dot{a}\over a} V_a  = 3 {\dot{a}\over a} c^2_a (V_a - V)
- {3\over 2 k}\left({\dot{a}\over a}\right)^2 (\Delta + 2 w \Pi)
\nonumber\\
& & +{k\over (1+w_a)}\left[ c_a^2 \Delta_{ca} + w_a \Gamma_a - {2\over 3}w_a
\Pi_a \right].
\end{eqnarray}
This equation can also be expressed in the following useful form:
\begin{eqnarray}
\label{delVdot}
(\dot{V}_a -\dot{V}) &+&  {\dot{a}\over a}(1-3c^2_a)(V_a-V)  =
+{k\over (1+w_a)}\left[ c_a^2 \Delta_{ca} + w_{a}\Gamma _{a}-
{2\over 3}w_a \Pi_a \right]
\nonumber\\
& & -{k\over (1+w)}\left[ c_s^2 \Delta + w \Gamma - {2\over 3}w \Pi \right].
\end{eqnarray}
For completeness, we also give the equations for the total density and velocity
perturbations which can be derived using eqs.
(\ref{totalfluid}-\ref{totalfluidend}, \ref{deldot}, and \ref{Vdot}):
\begin{equation}
\label{dtotdot}
\dot{\Delta} - 3 w{\dot{a}\over a}\Delta =  -k(1+w)V
-2 {\dot{a}\over a} w \Pi,
\end{equation}
and
\begin{eqnarray}
\label{Vtotdot}
\dot{V} &+&  {\dot{a}\over a} V  =
- {3\over 2 k}\left({\dot{a}\over a}\right)^2 (\Delta + 2 w \Pi)
\nonumber\\
& & +{k\over (1+w)}\left[ c_s^2 \Delta + w \Gamma - {2\over 3}w \Pi \right].
\end{eqnarray}

\subsection{Initial Values For $\sigma_\ell$}

   One can see from the equation (\ref{sigdot}) for the $\sigma_\ell$ that
only the $\ell=0$
and $\ell = 1$ modes are driven by gravitational coupling to the matter
fluctuations
in the universe.  If we look at the state of the free streaming particles when
the perturbation is well outside the horizon ($k\tau\ll 1$), we can see that
the higher moment equations (\ref{sigdot}) are solved by
%
%
\begin{equation}
\label{order}
    \sigma_\ell \sim (k\tau)^{\ell-1} \sigma_1,
\end{equation}
in this regime.  However, as we will discuss in section 4 on initial
conditions, $\sigma_0 \sim k\tau \sigma_1$, so in fact the
$\ell=2$ mode $\sigma_2 \sim \sigma_0$ and must be included in the definition
of initial conditions.  We can neglect all $\ell\geq 3$ moments, however.

  If we start the
evolution at $\tau_i$ when all scales of interest are outside the horizon
$k\tau_i\ll 1$, then we only need to specify the first three moments.  From
equation (\ref{Fdot}) for ${\cal F}$, we see ${\cal F}(\tau_i)$ ought to
have an initial form like (see also Peebles, 1973)
\begin{equation}
\label{Finit}
{\cal F} = {df^0\over dv}\left[ -{v\over 3} A + i q \mu B  + {v\over 3} C
{1\over 2}(3 \mu^2-1)\right]
\end{equation}
where the factors $v/3$ and $q$ have been chosen for the simplest
interpretation of $A$, $B$, and $C$.

{}From the definitions of $\Delta_{ca}$, $V_a$, and $\Pi_a$ we can see that
\begin{equation}
A={\Delta_{ca}\over 1+ w_a}(\tau_i)
\end{equation}
\begin{equation}
B= V_a(\tau_i)
\end{equation}
and
\begin{equation}
C= {5\over 3} {w_a\over c_a^2} {\Pi_a(\tau_i)\over 1+w_a}
\end{equation}

   Using the definition of $\cal F$ in eq. (\ref{Ftosig}) we can now make our
identification of the starting values for the $\sigma_\ell$:
\begin{eqnarray}
\label{siginit}
\sigma_0  &=&  -{4\pi\over 3} v {df\over dv} {\Delta_{ca}\over 1+w_a} \\
\sigma_1 &=& {4\pi\over 3} q {df\over dv} V_{a} \\
\sigma_2  &=& -{4\pi\over 45} v {df\over dv} {w_a\over c_a^2} {\Pi_{a}\over
1+w_a},\\
\sigma_\ell &=& 0\ \ \ \ {\rm for}\ \ell \ge 3
\end{eqnarray}
where all variables are to be evaluated at the initial time $\tau_i$.

   The initial conditions needed for setting up the fluid variables
$\Delta_{ca}$, $V_a$, and $\Pi_a$ will be discussed in section \ref{ics}.
   We now have all the ingredients necessary for solving the coupled set of
equations for an arbitrary composition of different matter components for
the universe.

\subsection{ Why ``Quasi-Newtonian", and the Physical Interpretation of the
Equations}
\label{meaning}

   The main reason for the calling our formalism ``quasi-Newtonian" is that
the equation for the gravitational potential has exactly the Newtonian form.
The gauge--invariant gravitational potential $\Phi$ and our
density perturbation variable are related via Poisson's equation, which in
Fourier space is:
\begin{equation}
\label{Poisson}
 {k^2\over a^2} \Phi = 4 \pi G \rho \Delta
\end{equation}
With other choices for a gauge-invariant density perturbation, this equation
only holds on sub-horizon scales.  Since the initial
conditions are usually specified when the perturbation scales are larger than
the horizon, it is convenient for the variables to behave in a simple intuitive
manner outside the horizon.  The equations in these variables then become
easily interpretable in terms of physics (of the sub-horizon behavior) in this
formulation.

  The Newtonian resemblance of this formalism goes deeper than the Poisson
relation.  The equations themselves have the characteristic that these are very
nearly the equations Newton would have written for an expanding gas acting
under the influence of its own gravity.  Note that in an expanding gas
(in $\Omega=1$ models), one can redefine the Newtonian potential in such
a way that equation (\ref{Poisson}) holds
(see, e.g., Peebles, 1980; section II.7).  The ubiquitous expression
\begin{equation}
\label{pressr2}
{k\over a (1+w)}\left[ c_s^2 \Delta + w \Gamma - {2\over 3}w \Pi \right]
\end{equation}
can be interpreted in a simple Newtonian way.  The first two terms are simply
the acceleration (in the rest frame of the matter) due to the pressure gradient
force, and the last term is the acceleration due to the divergence of the
anisotropic part of the stress tensor.  In the case of a
universe filled with cold matter, the equations for
$\dot{V}$ and $\dot{\Delta}$ have exactly the same form as the Newtonian
equations in expanding coordinates.  This is why we refer to our treatment as
the ``quasi--Newtonian" formulation.

   This formulation is closest to working in the gauge known as the
``velocity--orthogonal isotropic gauge" (Kodama \& Sasaki, 1984) in which there
is no residual gauge freedom.  We will elaborate more on the physical meaning
of these variables and the nature of this gauge in a future publication (Lyth
\& Schaefer, in preparation).   In both the $\Delta_{ca}$ and $V_a-V$
equations the term (\ref{pressr2}) appears.  This term is in turn equal to
\begin{equation}
\label{potderiv}
{1\over (1+w)}{\dot{a} \over a} \left[ c_s^2 \Delta + w \Gamma
- {2\over 3}w \Pi \right] =
-\left(\Phi - {1\over k}{\dot{a}\over a}V\right)^{\bf \cdot}
\end{equation}
which is effectively zero in a cold matter dominated universe.   The quantity
$\Phi - {1\over k}{\dot{a}\over a}V$ is the perturbation of the
expansion rate due to the matter perturbations.

   Note that all of our evolution equations depend only on the
perturbations in the distribution of particles.
We do not need an explicit solution of
the behavior of the metric perturbation as in the gauge dependent approach
(Bond \& Szalay, 1983; Ma \& Bertschinger, 1995).   We also do not need an
explicit solution of the potential equation (for $\Phi$) as in Durrer \&
Straumann, (1988) or Stompor, (1994).  All of the background cosmological
information is carried in the equation for the expansion rate.  To solve these
equations then, the only remaining piece of information we need is
the initial conditions.  Before getting to these, we would like to derive a set
of fluid equations for purely relativistic particles, as this would represent a
great simplification over the equations in this section (for this component).

\section{Equations for Collisionless Relativistic Particles}
\label{relcoll}

   If the particles are effectively massless over the range of time of interest
in the universe, so that $v=q$ is at least a good approximation, we can
integrate the
equations for the $\sigma_\ell$ to get equations for fluid variables.  In this
case $\Gamma_a$ will be zero.

   We multiply the equations for $\sigma_\ell$ by $v^3$, set $q=v$ and then
integrate over $v$ to get the fluid variables.  The first three fluid
variables $\Delta_{ca}$,
$V_a$, and $\Pi_a$ have been defined previously.  We repeat them again here for
the case of relativistic particles for completeness.  Using the subscript $r$
for relativistic fluids:
\begin{equation}
\label{del_r}
\Delta_{cr} =  {T_r^4\over \rho_r}\int_0^\infty dv v^3 \sigma_0
\end{equation}
\begin{equation}
\label{V_r}
V_r = -{3T_r^4\over 4\rho_r}\int_0^\infty dv v^3\sigma_1
\end{equation}
\begin{equation}
\label{Pi_r}
w_r \Pi_r = {T_r^4\over \rho_r}\int_0^\infty dv v^3\sigma_2
\end{equation}
\begin{equation}
\Gamma_r = 0
\end{equation}
and we also have higher moments\footnote{Note that although the same symbol
is used in Schaefer (1991), these
$\Pi^{(\ell)}$ are defined to be a factor of 3 larger than in Schaefer
(1991).}  which we call $\Pi^{(\ell)}$ corresponding to $\sigma_\ell$
for $\ell\geq 3$:
\begin{equation}
\label{PIl_r}
w_r \Pi_r^{(\ell)} = {T_r^4\over \rho_r}\int_0^\infty dv v^3\sigma_\ell;\ \
l\geq 3
\end{equation}

   If one were observing the radiation pattern from the origin of this
coordinate system at time $\tau$, the $\Pi^{(\ell)}$ would be the amplitudes of
the coefficients of a multipole moment expansion of that pattern.

The lowest moments equations are coupled by gravity to the other
matter in the universe.  The equations for these are

\begin{equation}
\label{deldot_r}
\dot{\Delta}_{cr} + {4\over 3} k V_r - {4\over 1+w} {\dot{a}\over a}
\left[ c_s^2 \Delta + w \Gamma - {2\over 3} w \Pi \right] = 0
\end{equation}

\begin{equation}
\label{Vdot_r}
\dot{V}_r -{k\over 4}(\Delta_{cr} - {2\over 3}\Pi_r) + {3\over 2k}
\left( {\dot{a}\over a} \right)^2[\Delta + 2 w \Pi] +{\dot{a}\over a} V = 0
\end{equation}

\begin{equation}
\label{Pidot_r}
\dot{\Pi}_r - {k\over 5}(8 V_r + 3\Pi_r^{(3)}) = 0
\end{equation}

\begin{equation}
\label{Pildot_r}
\dot{\Pi}_r^{(\ell)} + {k\over 2\ell +1}\left[ \ell \Pi_r^{(\ell-1)} -
(\ell+1)\Pi_r^{(\ell+1)}\right] = 0
\end{equation}

These moments are used for describing
the massless neutrino and decoupled photon fluctuations in the universe.

\section{Initial Conditions: Solutions for Super--Horizon Perturbations In
the Radiation Dominated Epoch}
\label{ics}

To numerically integrate structure formation models, one requires an accurate
set of solutions which may be used as initial conditions.  To  that end, we
consider solutions from both of the ``independent" types of  perturbations,
adiabatic and isocurvature, on  super-horizon scales.  Adiabatic fluctuations
typically arise from  inflationary models and isocurvature modes often result
from phase  transitions in the early universe.  In each type we will solve for
the behavior of the growing mode of the perturbations well outside the horizon,
deep within  the radiation dominated era.  These solutions will be the most
relevant for  starting numerical calculations.  We will start first with the
simpler adiabatic fluctuations.

\subsection{Adiabatic Perturbations}

In an adiabatic perturbation, the density fluctuations  $\delta \rho_a$ in
in the different components ($a$, $b$, {\it etc.}) have amplitudes which
are related by
\begin{equation}
\label{adiabatphys}
{\delta \rho_a \over (1+w_a) \rho_a} = {\delta \rho_b \over (1+w_b) \rho_b} =
\cdots = {\delta \rho \over (1+w) \rho}
\end{equation}
in order that the entropy per particle does not change.  With our choice of
gauge invariant density fluctuation variable, the same condition applies
{\em even when we are outside the horizon}.  Thus the adiabatic initial
conditions are exactly as expected intuitively:
%
%
\begin{equation}
\label{adiabatdelt}
{\Delta _{ca} \over (1+w_{a})} = {\Delta _{cb} \over (1+w_{b})} = \cdots =
{\Delta \over (1+w)},
\end{equation}
for all components.   We now proceed to identify the growing mode solutions
since these are the only ones that will survive long after their formation.
It will be assumed that we are beginning our simulation deep enough in the
radiation dominated epoch that the temperature is much larger than the
mass of any hot components.   Look first at eq.\
(\ref{delVdot}). Because we are considering super--horizon scales, we can
ignore all terms of order $(k\tau) \Delta $, an assumption which may be
validated self--consistently.  It is obvious that $V_{a} = V$ is a consistent
solution for all components:
%
%
\begin{equation}
\label{adiabatvel}
V_{a} = V_b = \cdots = V.
\end{equation}
Now consider eq.\
(\ref{deldot}), the equation for the growth of $\Delta_{ca}$. For each
component, $\Gamma_{a}$ and $(c^{2}_{a}-w_{a})$ are both negligible, and,
since the temperature is much larger than the mass of any hot component,
the factor $1+w_a$ is a constant for each component as well. So, eq.\
(\ref{deldot}) shows that the evolution of $\Delta _{ca}/(1+w_{a})$ is the
same for each component, preserving the adiabatic initial conditions. We
are left with the greatly simplified problem of solving only one set of
component equations. One can see from the equations for the evolution of
the collisionless particles that we will need to retain the first three
moments $(\Delta , V$ and $\Pi )$. For the photons, due to their strong
interaction with the baryons, and for the cold components, we only retain
the first two moments. The total anisotropic pressure, $\Pi $, which
appears in the evolution equations for $\Delta $ and $V$ can be found by
solving for the anisotropic pressure of the collisionless components, $\Pi
_{cr}$. The total $\Pi$ is related to the the collisionless components by
the ratio of the collisionless to total pressure, {\it i.e.}~$\Pi =
(p_{cr}/p)\Pi _{cr}$ where $p_{cr} $ is the pressure contribution from the
collisionless components and $p$ is the total pressure. It is left as a
exercise for the reader to verify that the solutions for curvature
perturbations have the following dependence on $\tau$:
%
%
\begin{eqnarray}
\label{curvsoln}
\Delta (k,\tau)  &=& \Delta_H \hat\alpha (\vec{k}) k^2 \tau^{2} \\
V(k,\tau)  &=& -{3 \over 4} {\Delta(k,\tau) \over k\tau } \left[ 1+ {2 p_{cr}
\over 5 p} \right]^{-1} \\
\Pi(k,\tau)  &=& - {3\over 5} \Delta (k,\tau) \left[ 1+ {2 p_{cr}
\over 5 p} \right]^{-1},
\end{eqnarray}
where $\hat{\alpha}(\vec{k})$ is a stochastic variable which carries
information about the probability and spectrum of the initial fluctuations,
and $\Delta_H$ is the amplitude of density fluctuation at {\sl Horizon
crossing}, {\it i.e.}, when $k\tau=1$.  $\hat{\alpha}(\vec{k})$ is usually
assumed to be a Gaussian random variable [{\it i.e.} with random phase -- see
Bardeen, {\it et al.}, 1986], and
was inspired by the formalism of quantum field theoretic description of the
quantum fluctuations during inflation (Guth \& Pi, 1982; Abbott \& Wise,
1984). When averaged over an ensemble of universes (or equivalently
over many horizon volumes), $\hat{\alpha}(\vec{k})$ has a mean of zero and a
variance defined by
\begin{equation}
\label{alphavar}
\langle \hat{\alpha}^\star(\vec{k}) \hat{\alpha}(\vec{k} ^{'}) \rangle =
(2\pi)^6 k^{n-4} \delta^3(\vec{k}-\vec{k} ^{'})
\end{equation}
where the brackets denote the ensemble average and the $\star$ signifies the
complex conjugate.  $n$ is the usual spectral index, and $n=1$ defines the
scale free (Harrison-Zeldovich) spectrum.  The factors of $2\pi$, which are not
in Abbott \& Wise, (1984), appear here
because we have associated the $1/(2\pi)^3$ factor with the wavenumber
integral of the Fourier transform, instead of with the space integral
transform.

Note that $\Delta_H$ above differs from the
more common specification of the amplitude at {\sl Hubble crossing} when the
wavenumber is equal to the Hubble length ($ka/\dot{a}=1$).  Thus the
amplitude $\Delta_H$ is a factor of 4 larger than the parameter
$\epsilon_H$ of Abbott \& Wise (1984).  We have chosen to use the Horizon
crossing specification so that $\Delta_H$ has the same definition in the
radiation and matter dominated epochs.  Using the more common Hubble crossing
definition leads to the use of different amplitudes for radiation and matter
dominated epochs (see {\it e.g.} Bardeen, Steinhardt, \& Turner, 1983).
In the strongly radiation dominated phase $\tau$ is related to the scale
factor by
\begin{equation}
\label{atotau}
   \tau \simeq {a \over H_0} \sqrt{ {\rho_m^0 \over \rho_r^0}}
 = 4.66 \times 10^5 a,
\end{equation}
where $H_0$ is the present Hubble constant, and $\rho_m^0$ and $\rho_r^0$ are
the present day matter and radiation (as if there were 3 types of relativistic
neutrinos) energy densities.
The above number assumes the COBE central value for the
current photon temperature $T= 2.726$ K (Mather, 1994).

   The power spectrum $P(k)$ of adiabatic perturbations is the Fourier
transform of the two point density correlation function.  Using our notation it
can be simply computed from the following definition:
\begin{equation}
\label{Pkdef}
\langle \Delta^\star(\vec{k},\tau) \Delta(\vec{k} ^{'},\tau) \rangle \equiv
(2\pi)^3 P(k,\tau) \delta^3(\vec{k}-\vec{k} ^{'}).
\end{equation}
Our initial power spectrum at $\tau_i$ is then
\begin{equation}
\label{Pk}
P(k,\tau_i) = (2\pi)^3 \Delta_H^2 k^n \tau_i^4
\end{equation}
The evolution equations in the previous sections can then be used to get the
present day $\Delta(k,\tau_0)$ and {\it via} eq (\ref{Pkdef}) the present
day power spectrum.

\subsection{Isocurvature Perturbations}

Isocurvature perturbations are distinctly different from the curvature type in
that the variation is not in local energy density, but rather in local
content.  We specify the initial conditions as perturbations which violate
eq.\ (\ref{adiabatdelt}), in such a way as the total perturbation
vanishes ($\Delta = 0$).  The relevant variable for parameterizing an
isocurvature perturbation is then naturally
\begin{equation}
\label{sab}
S_{ab} = {\Delta_{ca} \over 1+w_a} - {\Delta_{cb} \over 1+w_b}
\end{equation}
which measures the degree of non--adiabaticity.  Note that
in the adiabatic case there was only one set of initial conditions; here we see
that isocurvature intial conditions allow for many different possibilities.
Rather than discuss the system in general, it is more
instructive to consider a specific type of isocurvature perturbation and see
how to set up the isocurvature perturbations.  It is typical to consider the
type of perturbation where the perturbation in the matter is compensated by a
perturbation in the radiation.  For simplicity we will consider a case
where the matter is baryons and the radiation is only photons.  In this case
the density perturbations will have the following form:
\begin{eqnarray}
\label{exactics}
\Delta_{cb} &=& {S_{br} \over 1 + 3\rho_b/(4 \rho_r) } \\
\Delta_{cr} &=& -{ \rho_b \over \rho_r }{S_{br} \over 1 + 3\rho_b/(4 \rho_r) }.
\end{eqnarray}
However, we are considering the initial conditions deep in the radiation
dominated era, when we take $(\rho_b/\rho_r) \ll 1$.  When solving the
equations to find the growing mode, one must take care to keep only those terms
which are of the same order in small quantities.  For example, if we keep only
the zeroeth order quantities in $k\tau$ and $(\rho_b/\rho_r)$ in a universe
with only baryons and radiation, the solution would be
\begin{eqnarray}
\label{lowest order}
\Delta_{cb} &\approx& S_{br} \\
\Delta_{cr} &\approx& 0 \approx \Delta \\
V &\approx& 0 \approx V_r \approx V_b.
\end{eqnarray}
This is not accurate enough for numerical work, so we will find the leading
order dependence to insure that roundoff error does not induce an
unwanted adiabatic
perturbation.  In the specific case we are considering a small
adiabatic perturbation is actually induced because the two components do not
have the same sound speeds.  In fact, it is generally true that adiabatic and
isocurvature modes cannot be
completely separated when different components have different sound speeds.
(See, {\it e.g.}, Kodama \& Sasaki, 1984).

   We now solve for the leading order terms in the initial conditions where we
simply assume that $\Delta \ll \Delta_{ca}$.    To make our universe more
realistic, we will also include neutrinos which we assume to be distributed in
the same manner as the photons ($S_{br} = S_{b\nu}$).
\begin{eqnarray}
\label{dleading}
\Delta_{cr} &=& \Delta_{c\nu}= -{\rho_b \over \rho_r +\rho_\nu } \Delta_{cb}
\\ \nonumber
            &\approx& -{\rho_b \over \rho_r +\rho_\nu } S_{br}.
\end{eqnarray}
 Because $\Delta $ is
small, the term which drives the evolution of the fluctuations is the
entropy perturbation, $w\Gamma $, given in the relativistic epoch by
%
%
\begin{equation}
\label{Gammaiso}
\Gamma \approx -{\rho _{b} \over \rho_r +\rho_\nu } S_{br},
\end{equation}
The $\Gamma$ terms in equations (\ref{dtotdot}) and (\ref{Vtotdot}) then
drive $\Delta$ and $V$.
As was the case for adiabatic perturbations,
only the first three moments are significant. One can easily verify that
the solutions for isocurvature initial conditions are
%
%
\begin{eqnarray}
\label{isotsoln}
V(k,\tau)  &=& {1 \over 8} k\tau{\Gamma (k, \tau) }\left[ 1 +
{p_{cr} \over p} {2 \over 15} \right]^{-1} \\
\Pi(k,\tau)  &=& {8 \over 15} {p_{cr} \over p} {k \tau }V(k,\tau) \\
\label{isotsolnd}
\Delta(k,\tau)  &=& -{2 \over 3}{k\tau }V(k,\tau) - {1 \over 3}\Pi(k,\tau)
\end{eqnarray}

Furthermore, it can be shown
using eq.'s (\ref{deldot} \& \ref{Vdot}) that
%
%
\begin{eqnarray}
\label{isosoln}
\Delta_{cr} &=&  \Delta _{c\nu}  = \Gamma \\ \nonumber
V_{r}  &=& V_{\nu} = V_b = V.
\end{eqnarray}
As we will discuss in the next section, $V_b$ and $V_r$ are kept to be the same
by the strong scattering of photons and baryons ({\it via} electrons).
We can now clearly see
from eq.'s\ (\ref{isotsoln}--\ref{isotsolnd}) that our assumption that
$\Gamma $ induces total density perturbations $\Delta $ has been verified
self consistently.

   We have avoided introducing a random variable in the isocurvature case
because the fluctuations tend to come from non-gaussian sources.  The
probability
distribution function is then specific to the particular source of the
perturbations, and the accompanying formalism is perhaps best left unspecified.
In the case of Gaussian intial fluctuations, the
initial entropy perturbation $S_{ab}$ can be expressed using the stochastic
variables used in the adiabatic case: $S_{ab} = S \hat{\alpha}(\vec{k})$, where
$S$ is a constant amplitude of the intial perturbation.  The present day
power spectrum can be directly computed using eq. (\ref{Pkdef}).

\section{Equations for Photons: Baryon Coupling and Temperature Anisotropies}
\label{photonics}

  Observations of fluctuations in the temperature of the relic photons are an
important diagnostic tool for studying theories of structure formation.  The
techniques for relating the measured cosmic microwave anisotropies  to
theoretical models often use the predicted multipole moment spectrum.  We now
want to specify how these moments are calculated using our formalism to
facilitate comparisons of temperature anisotropies with estimates of the matter
fluctuation amplitudes.   Before doing this, however, we need to cover one
more topic which is necessary for calculating the evolution of the photon
distribution.

   Up to now we have only considered collisionless particles.   When matter is
ionized in the early universe, there is strong scattering between the photons
and electrons.  We discuss the treatment of this scattering in the next
sub--section.

\subsection{Photons Coupled to Baryons}

    The equations in section \ref{relcoll} work well for photons
when they are sufficiently decoupled from the baryons that they are
collisionless.  However in the early universe the temperature was high enough
to keep normal (baryonic) matter fully ionized.  The free electrons have a
large cross section for interacting with the photons and the baryons,
effectively coupling the baryons and photons tightly.
As time progresses, the temperature cools off to the point
where the electrons can combine with the protons (and nuclei) to form neutral
atoms, which have a small cross section.  We need equations to describe the
coupling of the photons and baryons.  The equations for this have been derived
elsewhere (Kodama \& Sasaki, 1984 - cf. Appendix E).  Here we will just give
the answer using the present formalism for the sake of completeness.  For a
more detailed explanation of some of the physical effects of on the cosmic
photon distribution, we refer the reader to Hu \& Sugiyama (1995), who use the
same density perturbation variables as in this work.

  The basic Liouville equation now has a collision term $C(f)$, which depends
on the distribution function $f$.  The equation becomes
\begin{equation}{\cal L}(f) = C(f).
\end{equation}
where $C(f)$ is the scattering functional which must be worked out for the
electromagnetic scattering of photons by electrons.  To get the fluid
equations, we must then integrate $C(f)$ over momentum in the same way that we
integrated the Boltzmann operator ${\cal L}$.  One can see that the
equations are just the same as the collisionless equations, but now with an
extra term added to describe the effect of scattering on the distribution.

The equations for the photons are as follows.  First, the equation for
$\Delta_{cr}$ is unchanged:
\begin{equation}
\label{deldot_p}
\dot{\Delta}_{cr} + {4\over 3} k V_r - {4\over 1+w} {\dot{a}\over a}
\left[ c_s^2 \Delta + w \Gamma - {2\over 3} w \Pi \right] = 0
\end{equation}

The fluid velocity equation (as well as the higher moment equations),
get scattering terms.
\begin{equation}
\label{Vdot_p}
\dot{V}_r -{k\over 4}(\Delta_{cr} - {2\over 3}\Pi_r) + {3\over 2k}
\left( {\dot{a}\over a} \right)^2[\Delta + 2 w \Pi] +{\dot{a}\over a} V =
an_e \sigma_T (V_b-V_r)
\end{equation}
where $n_e$ is the electron number density and
$\sigma_T= 6.6524 \times 10^{-25}$ cm$^2$ is the Thompson scattering cross
section.  The product $n_e \sigma_T$ is the photon collisional frequency.

The higher moments go as
\begin{equation}
\label{Pidot_p}
\dot{\Pi}_r - {k\over 5}(8 V_r + 3\Pi_r^{(3)}) = - {9\over 10} a n_e
\sigma_T \Pi_r
\end{equation}
and
\begin{equation}
\label{Pildot_p}
\dot{\Pi}_r^{(\ell)} + {k\over 2\ell +1}\left[ \ell \Pi_r^{(\ell-1)} -
(\ell+1)\Pi_r^{(\ell+1)}\right] = -a n_e\sigma_T \Pi_r^{(\ell)}
\end{equation}
We have included the $\cos^2\theta$ angular dependence in the differential
scattering cross section.  This is not explicitly evaluated in the formula in
Appendix E of Kodama \& Sasaki, but is included in their later work (Kodama
\& Sasaki, 1986; Gouda \& Sasaki, 1986).

The equations for the baryons can be obtained by knowing that the total
energy momentum tensor is conserved.  This means that the baryons behave
like cold matter with a
scattering term on the right hand side of the velocity equation.

\begin{equation}
\label{delVdot_b}
[\dot{V_b} - \dot{V}] +{\dot{a}\over a}[V_b-V]= -{k\over 1+w}
[c_s^2 \Delta + w\Gamma -{2\over 3}w\Pi] +{4\over3}{\rho_r\over \rho_b} a n_e
\sigma_T (V_r-V_b)
\end{equation}
while the density perturbation equation remains unchanged:
\begin{equation}
\label{deldot_b}
\dot{\Delta}_{cb} = -kV_b + {3\over 1+w} {\dot{a}\over a}[c_s^2 \Delta +
w\Gamma -{2\over 3}w\Pi]
\end{equation}

  To see the effect of the coupling induced by the scattering terms it is
instructive to look at the equations for the differences between the photon and
baryon fluid variables. This is done most conveniently in terms of the
variables $S_{br}$ and $V_{br}$ (Kodama \& Sasaki, 1984), where $S_{br}$ is the
relative entropy perturbation as in eq. (\ref{sab})
\begin{equation}
\label{sbr}
S_{br} = {\Delta_{cb}\over 1+w_b} - {\Delta_{cr}\over 1+w_r},
\end{equation}
and
\begin{equation}
V_{br} = V_b - V_r.
\end{equation}
After a little algebra we arrive at the equations
\begin{equation}
\label{Sbr}
\dot{S}_{br} = -k V_{br}
\end{equation}
and
\begin{equation}
\label{Vbr}
\dot{V_{br}} = -\left[ {\dot{a}\over a} +a n_e\sigma_T \left(1+ {4\over3}
{\rho_r\over \rho_b}\right)\right] V_{br}
- {\dot{a}\over a}(V_r-V) -{k\over 4}\left(\Delta_{cr}-{2\over3}\Pi_r \right)
\end{equation}
The key to understanding the behavior of the photon baryon system is
contained in the first term on the right hand side of eq. (\ref{Vbr}). If the
collision frequency ($n_e\sigma_T$) is much larger than the expansion rate
($\dot{a}/a^2$), then the two fluids are tightly coupled. Note that in this
regime small differences between the baryon and photon fluid velocities
will be strongly damped out. As long as $V_{br}$ is very small, then the
baryon density perturbation and the photon perturbation will be kept
proportional [in adiabatic perturbation cases, $\Delta_{cb} =
(3/4)\Delta_{cr}$].

   Through decoupling, it is numerically easier to integrate the equations for
the photons and $S_{br}$ and $V_{br}$.  Well before decoupling we can
obtain values for $V_{br}$ and $\Pi_r$ by assuming ``equilibrium"
conditions, that is, by using the value of $V_{br}$ and $\Pi_r$ for which
$\dot{V}_{br} = 0$ and $\dot{\Pi}_r=0$ in equations (\ref{Vbr} \&
\ref{Pidot_p}), respectively.  We have found that the tight coupling
limit (with these equilibrium solutions) is valid down to a temperature of
about 6000 K below which one should integrate the full equations.  After
decoupling, we switch from integrating equations for $S_{br}$ and $V_{br}$
in favor of those for the baryon perturbations $\Delta_{cb}$ and $V_b$.
$\Delta _{cb}$ ($V_b$) can be recovered using a linear combination of
$S_{br}$ ($V_{br}$) and $\Delta_{cr}$ ($V_r$).

\subsection{Evolution of Photon Perturbations in the Matter Dominated Epoch}

The equations introduced in the previous section are appropriate for
determining the evolution of the photon perturbations through the epoch of
recombination. However, to accurately track the photons until the present
epoch requires the retention of an inordinate number of terms in the
multipole expansion and is therefore numerically undesirable. Fortunately,
following recombination, it is possible to calculate an analytic solution
by making a few reasonable approximations. Most significantly, the
interactions between the photons and baryons can be ignored, and therefore
the collisionless Boltzmann equation may be solved. In the next
sub--section, we will estimate the errors made by using this
simplification.

Since we are primarily interested in calculating the
temperature fluctuation in the microwave background, it is convenient
notationally to write our equations in the variable ${\cal M} = \Delta
T/T$.  Given $\Delta T/T = (1/4)\Delta \rho /\rho $,
${\cal M}$ is related to the photon phase space distribution ${\cal F}$
by
%
%
\begin{equation}
\label{calM}
{\cal M} = {1\over 4} {T_r^4\over \rho_r}\int_{0}^{\infty } 4\pi dv v^{3}
{\cal F},
\end{equation}
where the multipole moments ${\cal M}_\ell$ are defined similar to the
$\sigma_\ell$ of ${\cal F}$ in
eq.\ (\ref{Ftosig}), {\it i.e.}
%
%
\begin{equation}
\label{M_ell}
{\cal M} = {1 \over 4\pi }\sum_{\ell =0}^{\infty
}(2\ell+1)i^{\ell}P_{\ell}(\mu ) {\cal M}_{\ell}.
\end{equation}
The variable $\cal{M}$ is more convenient to work with here, and it is
related to the variables of section \ref{relcoll} as
%
%
\begin{eqnarray}
\label{MtoPi}
(4\pi)^{-1} {\cal{M}}_0 & = & {1\over 4} \Delta_{cr} \\
(4\pi)^{-1} {\cal{M}}_1 & = & -{1\over 3} V_{r} \\
(4\pi)^{-1} {\cal{M}}_2 & = & {1\over 12} \Pi_{r} \\
(4\pi)^{-1} {\cal{M}}_\ell & = & {1\over 12} \Pi^{(\ell)}_{r}
\end{eqnarray}
Cosmic microwave anisotropy observations can be expressed in terms of
multipole moments derived from the spatially averaged temperature
autocorrelation function (looking along two direction vectors $\hat{x}_1$
and $\hat{x}_2$),
%
%
\begin{equation}
\label{autocorr}
\left \langle {\Delta T \over T}(\hat{x}_{1}) {\Delta T \over
T}(\hat{x}_{2}) \right \rangle = \sum_{}^{} Q^2_{\ell}P_{\ell}(\hat{x}_{1}\cdot
\hat{x}_{2}).
\end{equation}
One also sees this series in terms of the Abbott \& Wise (1984)
multipole moment coefficients $a_{\ell}$ which are related to the multipole
moments $Q_\ell$ by
%
%
\begin{equation}
\label{a_l}
a_{\ell}^2 = 4 \pi Q^{2}_{\ell}.
\end{equation}
Note that the $a_\ell^2$ used in Gorski, {\it et al.}, (1994) are a factor of
$1/(2\ell +1)$ times $a_\ell^2$ as defined above.  The theoretical multipole
moments, $Q^2_{\ell}$, are found by integrating over all the Fourier
components of ${\cal M}_{\ell}$, {\it i.e.}
%
%
\begin{equation}
\label{Ql}
Q^2_{\ell} = {1\over (2\pi)^3}\int_{0}^{\infty }dk k^{2}{2\ell+1 \over 4\pi }
\left | {\cal M} _{\ell} \right |^{2}.
\end{equation}
However,  while the $\cal{M}_{\ell}$'s are directly related to the observed
quantities, it is more convenient for solving the Boltzmann equation to
consider an alternate gauge invariant variable $\Theta $ (Kodama \& Sasaki,
1986), defined by
%
%
\begin{equation}
\label{Theta}
\Theta = {\cal M} -\left[ {1 \over k}{\dot{a} \over a}V - \Psi  \right],
\end{equation}
where $\Phi$ and $\Psi $ are the gauge-invariant potentials.  $\Psi$ is given
by
%
%
\begin{equation}
\label{Psi}
\Psi + \Phi = - 3\left( {\dot{a} \over ak} \right)^{2} w \Pi
\end{equation}
and $\Phi$ is defined in eq. (\ref{Poisson}).   Using this definition of
$\Theta $ along with the relation for the derivative of $\Phi $ expressed in
eq.\ (\ref{potderiv}), the collisionless Boltzmann equation for the photons
may be recast into
%
%
\begin{equation}
\label{Thetadot}
\dot{\Theta } +ik\mu \Theta = {\left[ \dot{\Psi} - \dot{\Phi}  \right]}.
\end{equation}
This equation can be solved by utilizing the integrating factor $\exp
(ik\mu \tau)$ producing the result
%
%
\begin{equation}
\label{Thetasoln}
\Theta = e^{-ik\mu (\tau - \tau_{i}) }\Theta ^{i}(\tau_i) +
\int_{\tau _{i}}^{\tau }d\tau '
e^{-ik\mu (\tau -\tau ')}{\left[ \dot{\Psi} - \dot{\Phi}  \right]}.
\end{equation}

We designate the first term in this solution as the intrinsic term since it
contains the intrinsic fluctuations as well as the Sachs--Wolfe effect. The
second term is called the integrated Sachs--Wolfe (ISW) effect, and we
will discuss it first. For the case of adiabatic perturbations in the
matter dominated epoch, the potentials $\Phi$ and $\Psi $ are effectively
constant over the length scales of interest.
One then can reasonably approximate the ISW effect by ignoring the
oscillatory part of the integral because most of the change in the
derivative occurs when $k(\tau - \tau') \ll 1$, {\it i.e.}\ the change
occurs before the oscillations become significant. When this condition is
violated, the SW effect is small relative to the intrinsic fluctuations and
will not affect the approximation. The resulting ISW effect is then given
by
%
%
\begin{equation}
\label{}
\left[ (\Phi -\Psi )-(\Phi -\Psi )_{i} \right]e^{-ik\mu (\Delta \tau)},
\end{equation}
with $\Delta \tau = \tau -\tau _{i}$.
The contribution to each multipole can be found by utilizing the
Legendre expansion of the exponential
%
%
\begin{equation}
\label{}
e^{ik\mu \Delta \tau } = \sum_{0}^{\infty } (2\ell +1)(-i)^{\ell}
j_{\ell}(k\Delta \tau ) P_{\ell}(\mu ),
\end{equation}
where $j_{\ell}$ is the spherical Bessel function.
When considering isocurvature perturbations,
the Sachs--Wolfe effect is small in proportion to the intrinsic
fluctuations, so one can ignore the ISW effect altogether.

Calculating the contribution from the intrinsic term is more difficult,
as it involves finding the moments of products of more than 2 Legendre
polynomials.
These have been calculated by Bond \& Efstathiou (1987) who showed that
%
%
\begin{equation}
\label{ThetaINT}
e^{-i\mu k\Delta \tau }\Theta_{l}^{i} = \sum_{l'}^{}\sum_{m=0}^
{{\rm min}(l,l')}\Theta
^{i}_{l'}(2l'+1) C_{l,l',m} (-1)^{l-m} j_{l+l'-2m}(k\Delta \tau),
\end{equation}
where the constant $C$ is
given by
%
%
\begin{equation}
\label{C}
C_{l,l',m} = \frac{c_{l'-m} c_{m}
c_{l-m}}{c_{l+l'-m}}\frac{2l+2l'-4m+1}{2l+2l'-2m+1},
\end{equation}
with
%
%
\begin{equation}
\label{a}
c_{n} = \frac{(2n-1)!!}{n!}.
\end{equation}
The advantage
to using this formulation should now be clear.  One need only retain the
number of terms significant after decoupling, which will be many fewer
than are needed for an accurate solution today, and calculate a single
summation rather than solve a large set of coupled differential equations.

\subsection{Adiabatic Sachs-Wolfe Multipole Moments}

   A useful approximation of the multipole moments $Q_\ell^2$ which can be
evaluated analytically is the case of the Sachs-Wolfe effect for
adiabatic perturbations.  The simplest versions of inflation predict Gaussian
adiabatic fluctuations so this case is of particular interest theoretically
(Abbott \& Wise, 1984, Abbott \& Schaefer, 1986; Fabbri, Lucchin, \&
Matarrese, 1987).  For this reason this approximation has been used
extensively in the analysis of COBE data (Wright, {\it et al.}, 1992; Smoot,
{\it et al.} 1992, Gorski, {\it et al.}, 1994; Bennett {\it et al.}, 1994).

  Our goal is to calculate the anisotropy in the photon distribution $\cal M$
at the present time $\tau_0$.  From eq. (\ref{Theta}),
\begin{equation}
\label{deltaTtheta}
{\cal M}|_{\tau_0} = \Theta |_{\tau_0} -\left[ \Psi - {1\over k} {\dot{a} \over
 a} V\right] |_{\tau_0}.
\end{equation}

   During the matter dominated era $a \approx \tau^2/\tau_0^2$ and
\begin{equation}
\label{cdmpotent}
\dot{\Phi} = 0 = \dot{\Psi},
\end{equation}
which is true even for if the matter is made of free streaming particles like
neutrinos as long as we stick to scales larger than the Jeans length $k\ll 0.02
(m_\nu/1\ {\rm eV})^{1/2}\ h$ Mpc$^{-1}$ (see {\it e.g.} Babu, {\it et al.},
1995), where $m_\nu$ is the neutrino mass.
In this case the solution of the Boltzmann equation in eq. (\ref{Thetasoln})
takes a particularly simple form:
\begin{equation}
\label{Thetacdmsoln}
\Theta |_{\tau_0} = e^{-ik\mu \Delta \tau} \Theta |_{\tau_d},
\end{equation}
where $\Delta \tau= \tau_0 - \tau_d$, and $\tau_d$ is the time when the photons
decouple from the baryons.

   On the largest scales, we can make the approximation ${\cal
M}|_{\tau_d}\approx 0$.  This is justified as follows.
${\cal M}$ is a linear combination of fluid variables $\Delta_{cr}$, $V_r$,
$\Pi_r$, and $\Pi_r^{(\ell)}$.  If $k\ll \tau_d^{-1}$ then to leading order,
${\cal M}|_{\tau_d}\propto (k\tau_d)^1$.  On the other hand, the potential
$\Psi \propto (k\tau_d)^0$, so ${\cal M}|_{\tau_d}\ll \Psi$.  This
approximation describes the redshifting of photons
by the gravitational potentials of matter perturbations, which is often
identified as the Sachs-Wolfe effect.  Using this approximation,
\begin{equation}
\label{deltaTpot}
{\cal M}|_{\tau_0} = \left[ \Psi - {1\over k} {\dot{a} \over a} V\right]
|_{\tau_d} e^{-ik\mu \Delta \tau} - \left[ \Psi - {1\over k}
{\dot{a} \over a} V\right] |_{\tau_0}.
\end{equation}

 During the matter dominated epoch,
\begin{equation}
\label{psisoln}
\Psi  = -6 \Delta_H \hat{\alpha}(\vec{k}),\ {\rm and}\ {1\over k}
{\dot{a}\over a} V = -4 \Delta_H \hat{\alpha}(\vec{k}).
\end{equation}
Using eq. (\ref{M_ell}), then we find for $\ell \geq 1$:
\begin{equation}
\label{M_elltoday}
{\cal M}_\ell |_{\tau_0} = (-1)^{\ell+1} 8 \pi \Delta_H \hat{\alpha}(\vec{k})
j_\ell(k\Delta \tau).
\end{equation}

For the multipole moment predictions we need the rms values of ${\cal M}_\ell$.
\begin{equation}
\label{M_rms}
\langle {\cal M}_\ell^\star (\vec{k}) {\cal M}_\ell (\vec{k} ^{'}) \rangle =
(2\pi)^3 |{\cal M}_\ell|^2 \delta(\vec{k} - \vec{k} ^{'})
\end{equation}
Using equation (\ref{alphavar}) and plugging the resulting value of
$|{\cal M}_\ell|^2 $ in eq. (\ref{Ql}) we get
\begin{equation}
\label{Q_rmsn}
Q_\ell^2 = 4 \pi^{3/2} \Delta_H^2 (2\ell +1) (\Delta \tau)^{1-n}
{\Gamma [3/2 - n/2] \over \Gamma[2 - n/2]} {\Gamma[ \ell + n/2 - 1/2]
\over \Gamma[ \ell -n/2 +5/2]}
\end{equation}
In the special case of the scale free Harrison-Zeldovich spectrum ($n=1$), this
reduces to
\begin{equation}
\label{Q_rmsn1}
Q_\ell^2 = 8 \pi \Delta_H^2 { (2\ell +1) \over \ell (\ell+1) }
\end{equation}
The above equation shows why the quantity $\ell (\ell +1) Q_\ell^2/(2\ell +1)$,
[called $\ell (\ell +1) C_\ell/(4 \pi)$ by some authors ({\it e.g.} Bond \&
Efstathiou, 1987; Crittenden, {\it et al.}, 1994; Hu \& Sugiyama, 1995; White,
Scott, \& Silk, 1994)] is often plotted versus
$\ell$.  If the Sachs Wolfe effect were the only
source of anisotropy, this quantity would correspond to a horizontal line.
Deviations from a straight line then illustrate the contributions from ${\cal
M}|_{\tau_d}$ and the integrated Sachs Wolfe effect.  We will plot the
results of a numerical calculation in section \ref{numerics}.

\subsection{The Effects of Relic Electrons and Reionization}

We have so far ignored the effects of scattering on our results for the
evolution of photon perturbations in the decoupled epoch, but it may have
significant consequences.  Recombination is not complete and there is always
present a small relic ionization.
The observations of Gunn \& Peterson (1965) indicate
reionization has occurred prior to a redshift of $z \approx 5$ (see
Giallongo {\it et al.} 1995 for recent results). We would now like to
consider these possibilities in our solutions and estimate their effects.

The Boltzmann equation with interaction terms can be written in the
following form:
%
%
\begin{equation}
\label{Thetadot_i}
\dot{\Theta } +ik\mu \Theta +R_{c}{\dot{a}\over a}\Theta = {\left[ \dot{\Psi} -
\dot{\Phi} \right]}+R_{c}{\dot{a}\over a}{\Theta_{0} \over 4\pi } -
R_{c}{\dot{a}\over a}V_{b}i\mu - R_{c}{\dot{a}\over a}{\Theta _{2}\over
8\pi }P_{2}(\mu),
\end{equation}
where $ R_{c} = a^{2}n_{e}\sigma _{T}/\dot{a} $ is the ratio of the
interaction rate to the Hubble expansion rate and $\Theta_\ell$ is related to
$\Theta$ the same way as ${\cal M}_\ell$ is related to ${\cal M}$, {\it i.e.},
through eq. (\ref{M_ell}). This leads to a modified integration factor which
includes a real component, $\exp\left({ik\mu \tau }+{\int_{a_{i}}^{a}da{R_{c}
\over a}}\right)$, that tends to wipe out fluctuations on
scales smaller than the scattering mean free path. Note that throughout
this section we implicitly assume that $a_{0} = 1$. Outside the horizon, as
can be seen from eq.'s (\ref{Sbr}~\&~\ref{Vbr}), the photon perturbations
grow like the dominant matter with scattering only affecting the growth of
the $\Pi $ term; scattering effects become important on the sub--horizon
scales. Well inside the horizon, the oscillating part of the integrating
factor tends to eliminate the effects of the inhomogeneous terms in the
Boltzmann equation, those appearing on the right--hand side of eq.\
(\ref{Thetadot_i}). As a first approximation, it is reasonable merely to
consider the homogeneous solution, that is to multiply the collisionless
solution by the damping factor $\exp\left({-\int_{a^{*}}^{a}da {R_{c}\over
a}} \right)$, where~$ a^{*}$ is defined as the greater of $a_{i}$ and
$a_{hc}$, defined as the scale factor for which $Ha_{hc}/k = 1$.

We now can use this approximate solution to estimate the effects of the
relic ionization on our collisionless results. In the matter dominated
epoch following recombination, we can write an expression for the damping
factor $R_{c} = .069 (1-Y_{p}) \Omega _{b}h \chi _{e} a^{-3/2} $, where
$Y_{p} = .23$ is the Helium mass fraction and $\chi _{e} $ is the Hydrogen
ionization fraction. As a particular example, consider a universe with $h =
1/2$ and $\Omega _{b} = .05$ which leaves a relic ionization of
approximately $\chi _{e} \approx 6 \times 10^{-4}$. Recombination occurs at
a redshift of $z_{rec} \approx 1250$, so if one uses the collisionless
solution beginning at $z = 300$ the damping factor is .997, meaning that at
most the error is less than .3\%. One makes only a negligible error by
neglecting the relic ionization.

If the universe is reionized, we can also estimate the damping effect.
Using the same model as the previous calculation, we assume that $\chi _{e}
= 1$ for redshifts less than $z \approx 5$, consistent with Giallongo \etal
(1995). We find a damping factor of .988 or a 1.2\% reduction.

\section{Numerical Solution of the Equations}
\label{numerics}

 We have implemented the equations presented here in two separate codes
which agree very well with each other and with other calculations
(Holtzman, 1989; Ma \& Bertschinger, 1995). The first code is a fixed
stepsize predictor corrector Haming type integrator which is extremely
accurate when integrating oscillatory equations. The second is an adaptive
stepsize Gear's method implicit differentiation solver which offers a good
balance between speed and accuracy. In the predictor corrector method, the
stepsize is chosen for each specific $k$ value, as the photon-baryon fluid
oscillates with period $2\pi/k$ in conformal time. This integrator was
reasonably fast. In the second method, we actually recast the equations
here as differential equations where the variable was the scale factor $a$.
During the matter dominated epoch $a$ is not a linear function of $\tau$,
so the stepsize needs to be changed continuously. This approach should of
course only be taken using with a method with an adaptive stepsize. If we
solve the perturbation equations as a function of $a$ we need only specify
the initial $a$, whereas with the fixed stepsize method we also have to
calculate the initial value of $\tau$ from eq. (\ref{atotau}).  Apart from
the differences mentioned above the two codes are otherwise identical.

 The equations for evolution of the fluids and the collisionless particles
are integrated explicitly to follow the evolution of density perturbations
in the early universe. For the collisionless particles we integrate the
$\sigma_\ell$ at 15 momenta values and then use Gauss-Laguerre integration to
get the associated fluid variables. We found the value of 15 momenta gave
reasonably accurate results (error in $\Delta_{c\nu}$ was around $1$ part in
$10^6$). The universe we have modeled contains 1 flavor
of massive neutrinos (collisionless particles), two flavors of massless
neutrinos, photons, baryons, and, in the adiabatic case, cold dark matter.
We treated the total fluid perturbations as
independent variables from the separate components which was especially
useful when dealing with isocurvature perturbations. The baryons and
photons are treated using a tight coupling approximation
until near recombination. The ionization fraction evolution
is then approximated by interpolation to a table of separate numerical
calculations (see Peebles, 1993). After $z= 300$, we no longer track the
evolution of the photons and neutrinos as individual components. Keeping
them requires the numerical evolution program to follow an exceedingly
large number of oscillations of the relativistic fluids after that time
when the wavelength is small ($\sim 1$ Mpc). The net effect is to lose the
influence of the anisotropic pressure in the equations governing the
evolution of the total perturbations. However, since the contribution goes
like $w\Pi $, it is not significant in the matter dominated epoch.

   We now present some results from the calculations for illustration.  First
we present the evolution of perturbations in an adiabatic cold plus hot dark
matter model (C+HDM) model with 5\% baryons and 25 \% hot dark matter (massive
neutrinos) and a Hubble parameter $h=0.5$.  This model was noted for
its attractive large scale structure properties long before the COBE results
were known (Schaefer, Shafi \& Stecker, 1989; see also Holtzman, 1989).  In
figure 1 we show the evolution
of the gauge--invariant fluid variables with time $\tau$ for $k=0.01\ h/$Mpc.
 In the three panels a), b) and c), we present the (component) perturbations
in density, velocity, anisotropic pressure, and neutrino entropy.  We also
present the total density and velocity perturbations.
The growth on this large length scale is not very exciting, as the massive
neutrinos ($m_\nu = 5.9\ $ eV) are effectively cold on this scale.  The
adiabatic conditions are preserved until the perturbation crosses the horizon
at $\tau = 1/k$.  The photon--baryon fluid begins to oscillate, but soon
experience decoupling at $\tau \sim 220$ Mpc.
The photons then, like the massless neutrinos,
simply free stream after that time.   The entropy in the neutrino distribution
grows but never becomes significant, especially when we consider that it only
enters the through the combination $w_\nu \Gamma_\nu$, where $w_\nu$ decreases
rapidly after the neutrinos become non--relativistic.  We stop following the
evolution of the relativistic components (photons and massless neutrinos) at
$\tau \sim 700$ Mpc, which corresponds to a redshift of $z\sim 300$.

In figure 2 we present similar results for $k=1\ h/$Mpc.  Here we start with
the same initial amplitude as in figure 1: $\Delta (\tau_i = 0.06\ {\rm
Mpc})=0.01$.  The evolution is more interesting.  First of all we see in
panel a) that the CDM component experiences growth which is damped (relative to
figure 1) by the effects of photons and neutrinos during the radiation
dominated epoch and by the free streaming of the massive neutrinos in the
matter dominated epoch.  The time of radiation matter equality is $\tau_{eq}
\approx 60\ $ Mpc.  The photon--baryon system undergoes acoustic oscillations
maintaining the adiabatic condition $\Delta_{cb} = (3/4)\Delta_{cr}$ until
decoupling.  Note that the velocity and density perturbations are 90$^\circ$
out of phase, as the particles flow in and out of the perturbation, the density
contrast must lag the velocity.  As the photons and baryons begin to decouple,
they undergo viscous (Silk) damping.  After decoupling, the baryons fall
into the potential wells of the CDM component.  The massive neutrinos free--
stream, but slowly settle into the CDM potential wells as the neutrino momenta
redshift.

    We next consider a case with isocurvature perturbations, the isocurvature
hot dark matter (IHDM) model.  We begin the model with the initial conditions
as
specified in eqs. (\ref{dleading} - \ref{isosoln}).  Again we plot the
results in figure 3 for $k= 0.01\ h$ Mpc$^{-1}$ and in figure 4 for
$k= 1\ h$ Mpc$^{-1}$.  Here there is no CDM component and with $h=0.5$ we have
$\Omega_\nu=1$ with a neutrino of mass $m_\nu = 23\ $ eV.  In panel c) we plot
the total entropy $\Gamma$, because of its role in driving the density
perturbation growth.  In figures 3 and 4
we see that $\Delta_{cb}$ is approximately constant until the induced adiabatic
curvature perturbation $\Delta$ becomes of the same order.  Outside the horizon
the neutrino and photon perturbations scale with the ratio of the matter
density
over the radiation density, while the total perturbation is down by a factor
$k^2\tau^2$ as in eq. \ref{isotsoln}.  The velocities and anisotropic pressure
perturbations scale with the amplitude of the induced adiabatic perturbations.

In figure 4, again we see that $\Delta_{cb}$ remains constant until the induced
adiabatic perturbation amplitude becomes comparable.  Now we also see that
despite the different amplitudes of $\Delta_{cb}$ and $\Delta_{cr}$ the
velocity perturbations oscillate together in the same way as in an adiabatic
perturbation.  We also see the viscous damping effect, and as in figure 2 the
massive neutrinos gradually fall into the potential wells set up by the
baryons.

 As an example of using our formalism for calculating temperature anisotropies,
we also calculate the multipole moments $Q_\ell^2$ for Gaussian $n=1$
spectra in an adiabatic CDM (figure 5) and an isocurvature HDM
universe (figure 6).  The multipole moments in figure 5 show nearly a
horizontal line for the low $\ell$, as expected from the Sachs-Wolfe
approximation of section 5.2.  We have normalized the moments to COBE, using
the quadrupole ($Q_2$) with the value $Q_{rms-PS} = 20\ \mu$ K.

\section{Conclusions}

  We have presented a simple formalism for following the evolution of density
perturbations in the early universe.  Our treatment includes the important case
of collisionless relic particles like massive neutrinos, which cannot be
accurately described as a fluid.  The density perturbations variables have
properties which make their evolution appear to be Newtonian with
an expanding space.  This makes it much easier to understand the physics of the
evolution, and we find this formalism to be very attractive for this feature
alone.  We have demonstrated the use of these equations
for transfer functions and temperature anisotropies in both adiabatic and
isocurvature models.

  We have shown examples here which demonstrate how the formalism works.  The
equations here can be easily modified to incorporate other cases, as we have
shown elsewhere (de Laix, Scherrer, \& Schaefer, 1995).  The initial
conditions are particularly easy to specify here (especially in isocurvature
models).  We have presented this new formalism as a working man's set of
equations which do not require a deep understanding of general relativity to
comprehend the behavior the perturbations under the action of gravity.

  Lastly, if one is interested in approximate transfer functions we have
presented a modified set of equations (in the Appendix) which can be integrated
extremely quickly on a computer.

{\bf Acknowledgements}  A. d. L. wishes to thank the DOE for support under the
grant DE-AC02-76ER01545.  R. S. wishes to acknowledge support for this
research under NASA grant NAG5-2646.

\section{References}

\begin{list}{    }{\listparindent -0.3in}

\item Abbott, L.F., \& Schaefer, R.K., 1986, ApJ, 308, 546

\item Abbott, L. F., \& Wise, M., 1984, ApJ, 282, L47

\item Babu, K. S., Schaefer, R.K., \& Shafi, Q., 1995, Bartol Preprint BA-95-19

\item Bardeen, J. M., 1980, Phys. Rev. D, 22, 1882

\item Bardeen, J. M., Bond, J.R., Kaiser, N., \& Szalay, A. S., 1986, ApJ,
{\bf 304}, 15

\item Bardeen, J. M., Steinhardt, P., \& Turner, M.S., 1983, Phys. Rev. D, 28,
679

\item Bennett, C. L., {\it et al.}, 1994, ApJ, 436, 423

\item Bond, J. R., \& Efstathiou, G., 1987, MNRAS, 226, 655

\item Bond, J. R., \& Szalay, A.S., 1983, ApJ, 274, 443

\item Crittenden, R. {\it et al.}, 1994, Phys. Rev. Lett., 71, 324

\item de Laix, A. A., Scherrer, R.J., \& Schaefer, R. K., 1995, ApJ, (in press)

\item Durrer, R., 1989, A\&A, 208, 1

\item Durrer, R., \& Straumann, N., 1988, Helv. Phys. Acta, 61, 1027

\item Ehlers, J., 1971, Proc. XLVII International School Enrico Fermi, ed B.K.
Sachs, (Amsterdam: North Holland)

\item Fabbri, R., Lucchin, F., \& Matarrese, S., 1987, ApJ, 315, 1

\item Giallongo, E., D'Odorico, S., Fontana, A., Savaglio, S., Cristiani, S.,
\& Molaro, P., 1995, astro-ph/9503005

\item Gorski, K., {\it et al.}, 1995, ApJ, 370, L5

\item Gouda, N. \& Sasaki, M., 1986, Prog. Theor. Phys., 76, 1016

\item Gunn, J. E., \& Peterson, B. A., 1965, Astrop. J., 142, 1663

\item Guth, A. \& Pi, S.--Y., 1982, Phys. Lett., 116B, 335

\item Holtzman, J. A., 1989, ApJS, 71, 1

\item Hu, W. \& Sugiyama, N., 1995, Phys. Rev. D, 51, 2599

\item Kodama, H., \& Sasaki, M., 1984, Prog. Theor. Phys., Suppl., 78, 1

\item Kodama, H., \& Sasaki, M., 1986, Int. J. Mod. Phys., A 1, 265

\item Lyth, D.H., \& Liddle, A. R., 1995, Astrop. Lett. \& Comm. (in press)

\item Lyth, D. H. \& Bruni, M., 1994, Phys. Lett., B323, 118

\item Lindquist, R. W., 1966, Ann. Phys., 37, 487

\item Ma, C.-P. \& Bertschinger, E., 1995, ApJ, (in press)

\item Mather, J.C. {\it et al.}, 1994, ApJ, 420, 439

\item Mukhanov, V.F., Feldman, H.A., \& Brandenberger, R.H. 1992, Phys. Rep.,
215, 203

\item Peebles, P. J. E., 1973, ApJ, 180, 1

\item Peebles, P. J. E., 1980, {\sl The Large Scale Structure of The Universe.}
(Princeton University Press: New Jersey)

\item Peebles, P. J. E., 1993, {\sl Principles of Physical Cosmology.}
(Princeton University Press: New Jersey)

\item Pogosyan, D. Yu., \& Starobinsky, A. A., 1994, MNRAS, 265, 507

\item Rebhan, A. K., \& Schwartz, D. J., 1994, Phys. Rev. D, 50, 2541

\item Schaefer, R. K., 1991, Int. J. Mod. Phys., 6, 2075

\item Schaefer, R. K., \& Shafi, Q., 1994, Phys. Rev. D, 49, 4990

\item Schaefer, R. K., Shafi, Q., \& Stecker, F.W., 1989, ApJ, 347, 575

\item Smoot, G.F., {\it et al.}, 1992, ApJ, 396, L1

\item Stewart, J. M., 1972, ApJ, 176, 323

\item Stompor, R., 1994, A\& A, 287, 693

\item Valdarnini, R., \& Bonometto, S. A., 1985, A\&A, 146, 235

\item White, M., Scott, D. \& Silk, J, 1994, AARA, {\bf 32}, 319

\item Wright, E.~L., {\it et al.}, 1992, ApJ, {\bf396}, L13

\end{list}

\section{Figure Captions}

\begin{enumerate}

\item  A plot of temporal evolution (in conformal time $\tau$) of
perturbations in a) density b)
velocity and c) anisotropic pressure for the various components and neutrino
entropy in an adiabatic cold + hot dark matter model (with
$\Omega_\nu=0.25$).  The perturbation wavenumber is $k=0.01\ h$ Mpc, and we are
using $h=0.5$.  We stop following the relativistic neutrino and photon
perturbations after a redshift of $\sim 300$.

\item  Similar to figure 1, using the same model, but for $k=1\ h$ Mpc.  Here
we can see the effects of viscous damping of the photon--baryon fluid, and
the gradual fall--in of the neutrinos (and the baryons after decoupling) into
the CDM potential wells.

\item Same as figure 1, but here we consider an isocurvature HDM model, which
has no CDM component.  The baryon density perturbation is compensated for by
the perturbation in the relativistic (photons and neutrinos) components.  The
rapid rise of the induced adiabatic perturbation $\Delta$ can be seen.

\item Same as figure 3, but for $k=1\ h$ Mpc.  As in figure 2 we can see the
various subhorizon evolutionary processes taking place.

\item  Multipole moments of the temperature anisotropy as a function of $\ell$
in an $n=1$, $h=0.5$, CDM universe, normalized to $Q_2 = 20\ \mu$ K$/T_0$.  If
the anisotropy were only due to the Sachs-Wolfe effect, the multipole moments
would fit a horizontal line up to the $\ell$ corresponding the to horizon size
at matter domination.

\item Same as figure 5, but for the isocurvature HDM universe. Here initial
perturbations in the photon distribution dominate the anisotropy.  The moments
are normalized to $Q_2 = 20\ \mu$ K$/T_0$.

\item  {\label{tranbar}} Comparison of transfer functions calculated using
the full treatment solid lines and the approximation described in the appendix
dotted lines for $\Omega_\nu = 1 - \Omega_{CDM} - \Omega_{b} = 0,$ 0.1, 0.2,
0.3, 0.4 and 0.5, using $h=0.5$ and adiabatic initial conditions.

\end{enumerate}

\appendix
\section{A Quick Method for Getting Transfer Functions}

     Setting up and running the codes to integrate the equations presented here
can be time consuming in terms of man-hours and CPU time.  If one only needs
an accuracy of about 10\% in the mass fluctuation, the procedure described
above can be replaced with a very simple set of equations.  These equations
have been discussed previously (Schaefer, 1991).  Here we will summarize that
method and compare it to the exact calculations of the current work.

     The method, based on the ``Grad" approximation of kinetic theory relies
on 2 simplifications.  First, for the collisionless particle components of
the universe, we ignore all the higher moments where ($\ell\ge 3$).  This means
we only consider $\Delta_{cr}$, $V_r$, and $\Pi_r$, and their equations, and
set $\Pi^{(\ell)}_r = 0$, for $\ell \ge3$.  This simplification, when used for
the massless relics, is well known
and has been exploited before ({\it e.g.,} Kodama \& Sasaki, 1986).
However, in order to use it effectively for the massive neutrinos, one needs a
second approximation.

The reason for this is that the equation for the massive neutrino $\Pi_\nu$
is problematic.  As described in Schaefer, (1991), one can derive the fluid
equations from the sigma equations.  However, the equations involve two
different momentum space integrals of each $\sigma_\ell$ (see eq.
\ref{sigdot}).  The corresponding fluid variables for these two momentum space
integrals for $\sigma_0$ are the density and pressure perturbations.  The
pressure perturbation shows up in the $\Gamma$ term in the equations presented
here.  If the particles are relativistic so that the energy and momentum are
equal, these two integrals are identical, and the problem disappears.
Schaefer (1991) advocated approximating the two integrals as
   \begin{equation}
\label{chiint}
c_\nu^2 \int dv q v^2 \sigma_\ell \approx  \int dv {v^4\over q} \sigma_\ell,\ \
(\ell=0,2).
\end{equation}

   Using this approximation, we find $\Gamma=0$ and we can derive an equation
for $\Pi$:
\begin{equation}
\label{schaeferpi}
\left({w_\nu\Pi_\nu\over c^2_\nu}\right)^{\bf \cdot} = -3 {\dot{a}\over
a}(c_\nu^2-w_\nu) \left( {w_\nu\Pi_\nu\over c^2_\nu}\right)
+ 3 k (1+w_\nu){2\over 5}V_\nu.
\end{equation}
The other two equations are
\begin{eqnarray}
\label{schaeferdv}
\left({\Delta_{c\nu}\over 1+w_\nu}\right)^{\bf \cdot} &=& -kV_\nu + {3\over
1+w}
{\dot{a}\over a}(c_s^2\Delta + w\Gamma -{2\over3}w\Pi),\\
\left(V_\nu - V\right)^{\bf \cdot} &=& -{k\over 1+w} (c_s^2\Delta +w\Gamma
-{2\over3}w\Pi)\nonumber \\
& & + {k\over 1+ w_\nu}(c_\nu^2 \Delta_{c\nu} -{2\over 3} w_\nu \Pi_\nu).
\end{eqnarray}

   These equations are certainly not adequate for calculating the temperature
anisotropy, because we have made a gross error in treating the photons.
However, the density transfer functions are reasonably accurate.  In Figure
(\ref{tranbar}) we show a comparison of transfer functions using the
approximations in this appendix and the full equations.  Over the interesting
range $k = 0 - 1.0\ h$/Mpc, the error is less than 10\%.

\end{document}